\documentclass{article}
\usepackage{graphicx}
\usepackage{glas}
\begin{document}

%Glasgow preprints
\begin{titlepage}{GLAS-PPE/2009-07}{15${\underline{\rm{th}}}$ May 2009}

\title{ScotGrid: Providing an Effective Distributed Tier-2 in the LHC Era}

\author{Ê 
Sam Skipsey $^1$, David Ambrose-Griffith $^2$, Greig Cowan $^3$ , Mike Kenyon $^1$, Orlando Richards $^3$, 
\\Phil Roffe $^2$, Graeme Stewart $^1$
%}
%\address{
\\
$^1$ University of Glasgow, University Avenue, Glasgow  G12 8QQ \\
%}
%\address{
$^2$ Department of Physics, Durham University, South Road, Durham DH1 3LE \\
%}
%\address{
$^3$ Department of Physics, University of Edinburgh, Edinburgh, EH9 3JZ
}

%\ead{s.skipsey@physics.gla.ac.uk}

\begin{abstract}
ScotGrid is a distributed Tier-2 centre in the UK with sites in Durham, Edinburgh and Glasgow. ScotGrid has undergone a huge expansion in hardware in anticipation of the LHC and now provides more than 4MSI2K and 500TB to the LHC VOs. Scaling up to this level of provision has brought many challenges to the Tier-2 and we show in this paper how we have adopted new methods of organising the centres, from fabric management and monitoring to remote management of sites to management and operational procedures, to meet these challenges. We describe how we have coped with different operational models at the sites, where Glagsow and Durham sites are managed ``in house" but resources at Edinburgh are managed as a central university resource. This required the adoption of a different fabric management model at Edinburgh and a special engagement with the cluster managers. Challenges arose from the different job models of local and grid submission that required special attention to resolve. We show how ScotGrid has successfully provided an infrastructure for ATLAS and LHCb Monte Carlo production. Special attention has been paid to ensuring that user analysis functions efficiently, which has required optimisation of local storage and networking to cope with the demands of user analysis. Finally, although these Tier-2 resources are pledged to the whole VO, we have established close links with our local physics user communities as being the best way to ensure that the Tier-2 functions effectively as a part of the LHC grid computing framework..

%preprints stuff
\vspace{0.5cm}
\begin{center}
{\em 17$^{\underline{\rm{th}}}$ International Conference on Computing in High Energy and Nuclear Physics}\\
{\em Prague, Czech Republic}
\end{center}
\end{abstract}

%preprints
\newpage
\end{titlepage}

\section{Introduction}
Unlike many other regions, the UK treats ``Tier-2" centres as primarily administrative groupings; UK-style ``distributed" Tier-2s consist of multiple sites, grouped by region (e.g. \emph{ScotGrid}\cite{scotgrid} for mostly Scottish sites, \emph{NorthGrid} for those in the north of England, \&c), but not abstracted at this level to present a single ``Tier-2 site" to the grid. Each component site is visible as a distinct cluster and storage, with its own Compute Element, Storage Element, Site-BDII and other components.
Each component site thus also has its own systems administration teams at the host location (mostly Universities), and has to abide by the host's requirements. Hierarchical control then passes through a ``Tier-2 Coordinator" who is responsible for sheparding the Tier-2 as a whole.
The ScotGrid Tier-2 thus consists of sites hosted by the Universities of Glasgow, Edinburgh and Durham, administered in the manner described above. Due to the lack of an established term for component sites of a distributed Tier-2, we will adopt the common terminology in the UK itself and refer to components as "Tier-2 sites" themselves, despite the ambiguity this introduces.

This paper will attempt to provide an overview of challenges and solutions we have encountered whilst establishing the Tier-2 and scaling it to provide the expected demand from the LHC.

\section{Fabric and People Management}
ScotGrid has adopted several common best practices for the administration of component sites.
For configuration and fabric management, the Glasgow and Durham sites have adopted cfengine\cite{cfengine} as an automation tool. Durham's configuration is derived from the original configuration scripts used by Glasgow, but omits low-level (non-cfengine) tools for imaging nodes and services.

All sites use the Ganglia\cite{ganglia} fabric monitoring tool to collect historical and real-time data from all nodes in their sites.
We have also configured a centralised nagios\cite{nagios} alerting system, monitoring Glasgow's systems directly, and Edinburgh and Durham via external tests (including SAM test monitoring provided by Chris Brew of RAL).

When failures or system problems are detected, a shared login server, hosted at Glasgow, allows sysadmins from any of the sites to log in with administrative rights to the other sites. This system is managed by ssh with public key security and host-based authentication, and is therefore fairly secure against external attacks.

Recently, we have also established a ``Virtual control room'', using the free, but proprietary Skype\cite{skype} VoIP and text messaging system. The control room, implemented as a ``group chatroom''. has persistent historical state for all users who are members, allowing the full history of conversations within the room to be retrieved by any user on login (including conversations that occurred during periods they were not present). We believe that this feature is not available as standard functionality in any other application.
The virtual control room has already proved of great value, enabling advice and technical assistance to be provided during periods of critical service downtime. As the Tier-2 is distributed, so is the expertise; a common control room allows this expertise to easily be shared in a way that community blogs and mailing lists can't replicate.

\subsection{Local services}
The gLite middleware has a natural hierarchical structure, such that information flows up from all services into a small number of ``top-level" services (``Top-level BDIIs'') with global information, and similarly control and job distribution flows out from a small number of ``Workload Management Services'' to all the compute resources.
 As such, it was originally envisaged that the Top-level BDII and WMS services would exist at the Tier-1 level. However, for historical reasons, and due to the unreliability of WMS services, ScotGrid manages its own copies of both services, which are used by all the sites in the Tier-2.
Top-level BDIIs are easy to administer, and do not require any complicated configuration; they are essentially ldap\cite{ldap} servers with some tweaks applied.

The glite WMS (technically two services: the WMS itself and the ``Logging and Bookkeeping'' service that maintains records) has developed a reputation for being hard to administer and maintain. The ScotGrid WMS at Glasgow was set up in order to reduce our dependance on the (then single) WMS instance at the Tier-1, removing the single point of failure this represented for job submission. Since then, the RAL Tier-1's WMS service has become somewhat more reliable, with the aid of extensive load-balancing; however, the WMS as a service is still the least stable part of the job control process, and is still a major cause of test failures against sites.  Other Tier-2s have begun installing their own ``local'' WMSes in order to reclaim control of these test failures, as it is psychologically stressful to experience failures outside of your influence.

In addition, Durham run their own WMS instance for use by local users and for training purposes. This is not enabled as a service for users outside of Durham currently; it is mentioned here to avoid confusion in the later discussion of Durham system implementation.

Glasgow also runs its own VOMS server, although not to duplicate central services at the Tier-1 or RAL. ScotGrid supports several local VOs, one of which will be discussed later in this paper, and the VOMS server is necessary to support user membership, authentication and authorisation control for those entities. Possessing a VOMS server gives us additional flexibility in supporting local users, especially those who are interested in merely ``trying out'' the grid, without wanting to invest in the overhead of creating their own VO.

\section{``Special cases''}
Whilst Glasgow's configuration is generally ``standard'' in terms of common practice at other UK Tier-2s, Durham and Edinburgh both have nonstandard aspects to their site configuration. Durham host their front-end services on virtual machines; Edinburgh's compute provision is via a share of a central university service, rather than a dedicated service for WLCG.

\subsection{Durham: Virtual Machines}
In December 2008, Durham upgraded their entire site with the aid of funding from . As part of this process, the front-end grid services were implemented as VMware\cite{vmware} virtual machines on a small number of very powerful physical hosts.

Two identical hosts, ``grid-vhost1.dur.scotgrid.ac.uk'' and ``grid-vhost2.dur.scotgrid.ac.uk'', configured with dual quad-core Intel Xeon E5420 processors, 16 Gb RAM and dual (channel-bonded) Gigabit ethernet. The services can therefore be virtualised easily in ``blocks'' of (1 core + 2 Gb RAM), allowing different services to be provisioned with proportionate shares of the total host. Using VMware Server 2.0 limits the size of a given virtual machine to two cores, which does present a practical limit on the power assignable to any given service. In many cases, this can be ameliorated by simply instantiating more than one VM supporting that service, and balancing load across them (this works for compute elements each exposing the same cluster, but storage elements cannot share storage and so cannot load balance in the same way).

The current configuration allocates one block each of grid-vhost1 to VMs for ganglia monitoring, cfengine and other install services, and the glite-mon service, two blocks to the VM hosting the first Compute Element. Two blocks plus the remaining 4Gb of RAM to the local Durham WMS service, as WMS services are extremely memory hungry. This leaves one core unallocatable to additional VMs as there is no unallocated memory to assign. 

Grid-vhost2 devotes one block each to the Site-level BDII service and the local resource management system (Torque\cite{torque}/Maui\cite{maui}), two blocks to a second CE service to spread load, and two cores and 2Gb ram to a storage element service using DPM. As will be shown later in the paper, this configuration does not provide sufficient compute power to the storage element, but load-balancing is non-trivial for DPM and other commonly implementations of the SRM protocol, so performance increases will have to be attempted via configuration tuning initially.

In all cases, the virtual machine files themselves are hosted on NFS filesystems exported from a master node, via dual (channel-bonded) Gigabit ethernet. The local filesystem is mounted on 2TB of fibre-channel RAID for performance and security. Interestingly, exporting this filesystem over NFS gave better performance than the local disks in the virtual machine hosts. 
As the virtual machine files are hosted on a central service, it is trivial to move services between the two virtual machine hosts, in cases where one host needs maintenance.

Outside of ScotGrid, the Oxford site (part of SouthGrid) has also installed front-end services as virtual machines. Considering the growing acceptance of virtualisation as a means of supporting multiple applications efficiently on the increasingly multicore architectures of modern CPUs, we see this becoming a mainstream configuration across the WLCG in the next five years.

\subsection{Edinburgh: Central Services}
Historically, Edinburgh's contribution to ScotGrid has been storage-heavy and compute-light. This was a legacy of the initial experiment of the ScotGrid configuration, where Glasgow was intended to host compute services and Edinburgh storage, with fast networking allowing these to act as mutually local resources. This turned out to not be feasible at the time, although NorduGrid\cite{nordugrid} has effectively proved that such a scheme can succeed, albeit with their own middleware stack rather than gLite.

As a result, the paucity of EdinburghÕs installed cluster resources lead to a decision, in 2006, to decommission the old site, replacing it with a new site with compute power provided by the then-planned central university compute resource, ECDF (the Edinburgh Compute and Data Facility).
As the ECDF is a shared university resource, its main loyalty must be to the local University users, rather than to the WLCG. This is a currently unusual, but increasingly common state of affairs, and requires some adjustment of expectations on both sides.

In particular, as local users canÕt be forced to use the grid middleware to submit their jobs (preferring direct acces), the normal installation process for worker nodes canÕt be performed as they pollute the local environment with non-standard libraries, services and cron jobs. Instead, the ``tarball'' distribution of the glite-WN is deployed, via the clusterÕs shared filesystem. However, even the tarball distribution requires cron jobs to be installed to maintain the freshness of CRL files for all supported Certificate Authorities; these are instead run from the Compute Element, which has access to the instance of the CA certificate directory on the shared filesystem. 

Services like the Compute and Storage Elements all run on services which are physically administered by the ECDF sysadmins, but software administered by a separate ``middleware teamÕÕ who were partly funded by GridPP to provide this service. As a result, the installation and configuration of these ``front-end nodes'' is broadly similar to that in other Tier-2 sites; the sole exception being the interface to the local resource management system, in this case Sun Grid Engine\cite{sge}. Again, the default configuration procedure for gLite middleware via YAIM involves configuration of the LRMS as well as the CEs that attach to it; this is highly undesirable in the case of a shared resource, as the queue configuration has usually already been arranged correctly. As a result, configuration of the CEs was performed partly by hand, and required some recoding of the jobmanager scripts provided by the gLite RPMs. 

Whilst the resulting system is functional, the necessity of not using YAIM to configure some elements of the compute provision (both CE and worker nodes) reduces the agility of the site in response to middleware updates. Despite this, Edinburgh/ECDF has managed to be a successful Tier-2 site, after all the pitfalls were worked through.

\section{WLCG/EGEE Use}
The configuration and management procedures mentioned in the previous section have enabled ScotGrid to provide significant, and reliable, resources to the WLCG VOs. The workload presented by these VOs can be divided into two main classifications: \emph{Monte-Carlo production}, the generation of simulated datasets for use in later analysis of detector data; and \emph{user analysis}, which comprises all activities involving ``realÕÕ data from the LHC itself. Production work by the LHCb and ATLAS VOs has been on-going on ScotGrid sites for several years, and is well-understood; User Analysis, conversely, has been necessarily limited by the delayed operation of the LHC itself. The LHCb compute model reserves Tier-1 sites for user analysis processes, and so we would only expect significant analysis activity from the ATLAS VO when the LHC begins operation.

\subsection{Production}
As Production work is well established, all VOs have detailed historical logs which can be accessed to investigate the relative effectivness of sites. We will briefly discuss some statistics from the LHCb and ATLAS VOs regarding recent Production work as concerns ScotGrid sites.

\subsubsection{LHCb}
\begin{figure}
\includegraphics[width=38pc]{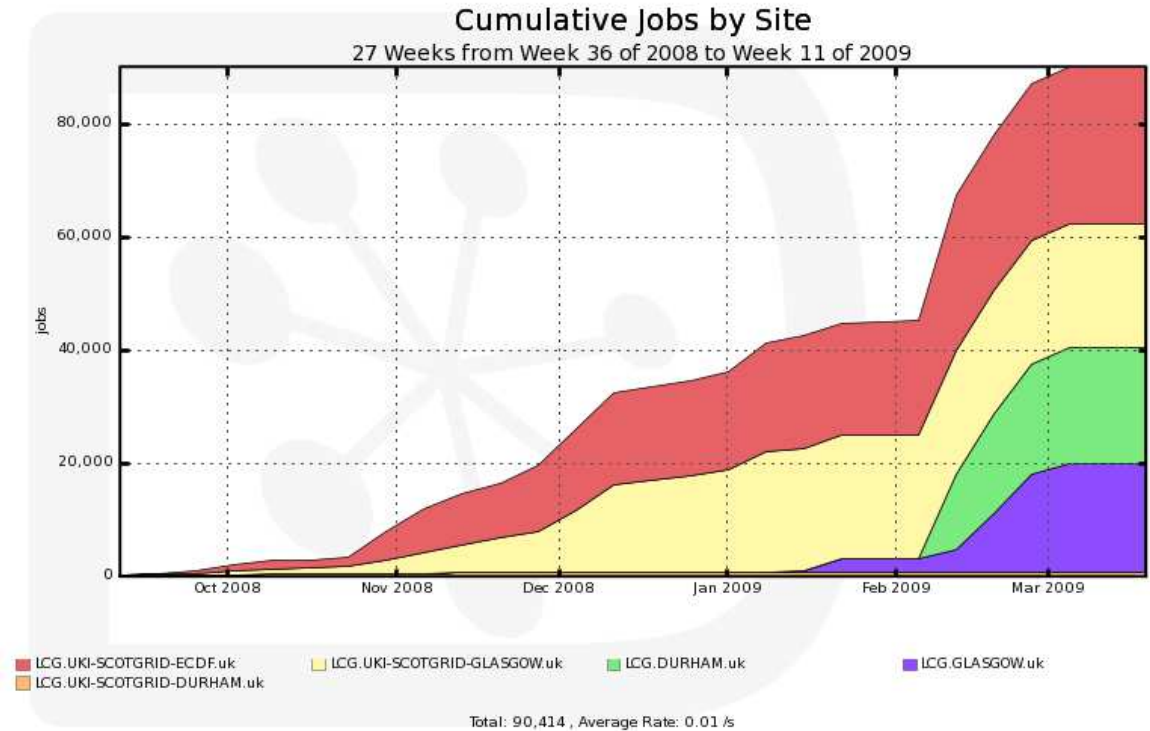}
\caption{Cumulative LHCb Production work on ScotGrid sites over a 6 month period ending in March 2009. Glasgow is represented by both blue {\it and} yellow areas due to a change in the internal names used to represent sites in the LHCb accounting software. Durham is represented by the green area and ECDF/Edinburgh by the red. \label{lhcbfigure}}
\end{figure}

Figure \ref{lhcbfigure} shows the number of successful LHCb production jobs accumulated by ScotGrid sites over a 6 month period ending in the 11th week of 2009. The point at which Durham's site became active is clearly visible as the rapidly increasing green area to the right-hand side of the plot. 

Between September 2008 and February 2009 inclusive, ScotGrid had processed almost 90,000 LHCb jobs; 20,000 of those arrived at Durham in February alone, although the ``bursty'' nature of LHCb production demand prevents us from concluding that Durham is as dramatically better than the other sites as this would initially seem. Indeed, almost half of the entire cumulative distribution occurs in the month of February, a reflection more on the VO's increased demand than on Tier-2 itself.

\subsubsection{ATLAS}
Table \ref{atlasprod1} shows the statistics for ATLAS Production jobs passing through ScotGrid sites during the middle of March 2009. As can be seen, Glasgow processed almost 18\% of the total ATLAS production workload in the sampled period. Considering that this data includes the Tier-1, which handled almost 50\% of the total, this is a particularly impressive statistic for ScotGrid. By comparison, Edinburgh is comparable to the average Tier-2 site, and Durham happens to look particularly unimpressive in this accounting because of on-going problems with ATLAS jobs not targeting their site. Even with these problems, Durham, in common with the other component sites, surpasses the average efficiencies for sites in the UK significantly.

A more recent sample of the state of ATLAS production, presented in table \ref{atlasprod2} shows that after DurhamÕs initial teething problems were fixed, it is fully capable of matching the performance of other UKI sites. Indeed, in this sampling, all of the ScotGrid sites surpass the average job (walltime) efficiency for the UK.
\begin{table}
\caption{\label{atlasprod1}ATLAS Production (March 09). Walltime figures are given in minutes,``Success'' and ``Failure'' values are total numbers of jobs.}
\begin{center}
\begin{tabular}{crrrcc}
%\br bold rule
\hline 
\hline
Site&Success&Failure&Walltime (success)&Efficiency&Walltime Efficiency\\
%\mr mid-weight rule
\hline
Glasgow&87376&16461&1859086649&84.1\%&96.4\%\\
Edinburgh&9103&1421&135756242&86.5\%&94.8\%\\
Durham&1427&113&504188&92.7\%&96.0\%\\
Total UK&489889&111219&$\sim$9$\times$10$^9$&81.5\%&88.9\%\\
%\br
\hline
\hline
\end{tabular}
\end{center}
\end{table}

\begin{table}
\caption{\label{atlasprod2}ATLAS Production (April 09). Taken from the first 21 days of April, reported in ATLAS Prodsys tool. Walltime figures are given in minutes,``Success'' and ``Failure'' values are total numbers of jobs.}
\begin{center}
\begin{tabular}{crrrcc}
%\br
\hline
\hline
Site&Success&Failure&Walltime (success)&Efficiency&Walltime Efficiency\\
%\mr
\hline
Glasgow&17970&629&379671909&96.6\%&99.6\%\\
Edinburgh&4019&581&48351886&87.4\%&97.7\%\\
Durham&6069&454&54405200&93.0\%&94.9\%\\
Total UK&294251&35696&$\sim$4$\times$10$^9$&89.2\%&92.9\%\\
%\br
\hline
\hline
\end{tabular}
\end{center}
\end{table}

\subsection{ATLAS User Analysis}
Although it has been suspected for some time that user analysis activities would present a different load pattern to production, until recently there has been surprisingly little work done in trying to quantify these differences. Indeed, if the LHC had begun operation on time, we would have been taking user analysis jobs without any solid knowledge of how they would perform.
The delayed operation of the LHC has allowed the ATLAS VO to begin a campaign of automated tests against all Tiers of the WLCG infrastructure in order to begin such quantification work. Leveraging the Ganga\cite{ganga} Python-scriptable  grid user interface framework, the HammerCloud\cite{hammercloud} tests submit large numbers of a ``typicalÕÕ ATLAS user analysis job to selected sites, allowing load representative of that from a functioning LHC to be presented on demand.

In general, it is found that the ATLAS analysis workload stresses storage elements much more significantly than production work can. Although the ATLAS VO is in the process of adapting its workflow to reduce such load, it is clear that provisioning of site infrastructure on the basis of production workflow has resulted in many sites, including those in ScotGrid, being significantly underpowered for optimal analysis performance. 

An analysis of the performance of Glasgow under HammerCloud tests, including optimisation work undertaken is presented our other paper in this proceedings\cite{dpmpaper}.
Durham has also begun testing the performance of their infrastructure, revealing similar problems to those at Glasgow before tuning was begun. Whilst 2 cores of an Intel Xeon processor are more than sufficient to support load on DPM services from production use, the extreme frequency of get requests generated by analysis jobs results in the node becoming saturated and effective performance bottlenecked. 

\section{Local Users}
As well as provision for WLCG/EGEE users, ScotGrid sites also have user communities local to each site. In this section, we will provide a brief overview of how these local users are provisioned and supported at Glasgow.

\subsection{Glasgow ``Tier-2.5"}
One of the largest groups of local users at Glasgow are the local particle physicists themselves. Whilst these users are all members of WLCG VOs and could simply submit jobs via the high-level submission frameworks, it is useful for them to be able to access other local resources when choosing to run jobs at the Glasgow site. This mode of operation, merging ``personal'' resources (often informally regarded as ``Tier 3'' of the WLCG) and Tier-2 compute provision is referred to as
``Tier-2.5'' provision within Glasgow.

The implementation mostly concerns tweaks to user mapping and access control lists, so that local users can be mapped from their certificate DN to their University of Glasgow username, rather than a generic pool account, whilst mapping their unix group to the same group as a generic VO member. This process involves some delicate manipulation of the LCMAPS credential mapping system: LCMAPS will only allow a given method of user mapping to be specified once in its configuration file; however, the Tier-2.5 mappings uses the same method (mappings from predefined list) as the default mapping method. The solution is to make a copy of the mapping plugin required, and call the copy for the Tier-2.5 mappings.

The resultant combination of user and group memberships allows Tier-2.5 users to access local University resources transparently, whilst also allowing them to access their WLCG-level resources and software. The provision of a local gLite User Interface service with transparent (ssh key) based login for local Physics users also helps to streamline access.

Similarly, the Glasgow site's storage elements allow local access to ATLAS datasets stored on them via the rfio and xroot protocols, through the use of their grid credentials whilst mapped to their local accounts. As the capacity of the Glasgow storage system is significant, this allows large data collections to be easily cached locally by researchers without the need for external storage.

\subsection{Glasgow NanoCMOS}
The other significant user community at Glasgow are represented by the ``NanoCMOS'' VO, which represents electrical engineers engaged in simulation of CMOS devices at multiple levels of detail (from atomistic simulations to simulations of whole processors). As the only supported means of submitting to the Glasgow cluster is via the gLite interfaces, local NanoCMOS members currently use Ganga to submit jobs to the Glasgow WMS instances. 

The typical NanoCMOS workload involves parameter sweeps across a single simulation, in order to generate characterisation of I/V curves and other data. This means that a single user will submit hundreds of subjobs to the WMS, which has challenged the scalability of the existing infrastructure. (Compare this to the recommendation that batches of jobs submitted to a gLite WMS should be of order 50 to 100 subjobs). Even the lcg-CE service has scalability problems in receiving on the order of 1000 jobs in a small time period; whilst it can manage thousands of concurrent runningjobs, the load on the CE is significantly higher during job submission and job completion, leading to component processes sometimes zombifying under the stress of massively bulky submission. Educating NanoCMOS users to stagger job submissions mitigates their effect on the system load, but this is an on-going problem. We are currently exploring other CE solutions, both the CREAM-CE and the NorduGrid ARC-CE, in order to examine their performance under similar load.
   
\section{Conclusions}
This paper has presented an overview of the configuration and management decisions made at the ScotGrid distributed Tier-2. We have discussed the ways in which two of our component sites deviate from the ``traditional'' configuration of a WLCG site; both of which are likely to become mainstream given time.

In general, we have found that means of enhancing communication between the component sites are the most effective things to implement. The ``virtual control room'', in particular, has been of signal benefit to all of the sites in organising effort and transferring expertise in real-time.
There is also an apparently unavoidable tension between the demands of WLCG use and those of local users, both for a traditional site with wholly-gLite mediated interfaces like Glasgow, and for a highly non-traditional site with a bias towards low-level access like Edinburgh/ECDF.
LHCb and ATLAS statistics demonstrate that we are very successful Tier-2 from the perspective of Monte-Carlo production, which makes the difficulties we have had with simulated User Analysis load particularly notable (although much of this discussion is outside of the scope of this paper). It is likely that this will be our most significant issue in the coming year, although work is continuing to address this challenge.

\section*{Acknowledgements}
This work was supported by the GridPP project, funded by the UK Science and Technologies Facilities Council.

\end{document}